\documentstyle[psfig]{l-aa}             

\begin{document}

\def\diff{\partial} 
\def\d{ {\rm d} }   
\def\vec#1{\bf #1}  

\def\n{\noindent}
\let\speciali =\n
\def\ol{\overline}
\let\specialii =\ol
\def\O{\Omega}
\let\specialiii =\O
\def\wt{\widetilde}
\let\specialiv =\wt
\def\wh{\widehat}
\let\specialv =\wh
\def\f{f(\theta)}
\let\specialvi =\f
\def\k{\rho \chi}
\let\specialvii =\k
\def\en{\rho \varepsilon}
\let\specialviii =\en


\newcommand{\beq}{\begin{equation}}
\newcommand{\beqa}{\begin{eqnarray*}}
\newcommand{\beqan}{\begin{eqnarray}}
\newcommand{\greq}{\begin{equation}\left\{ \begin{array}{l}}
\newcommand{\eeq}{\end{equation}} 
\newcommand{\eeqa}{\end{eqnarray*}}
\newcommand{\eeqan}{\end{eqnarray}}
\newcommand{\lp}{ \left(}
\newcommand{\rp}{ \right)}
\newcommand{\lc}{ \left[}
\newcommand{\rc}{ \right]}
\newcommand{\eps}{\varepsilon}
\newcommand{\intsur}{ \int_{(S)}\! }
\newcommand{\intvol}{ \int_{(V)}\! }
\newcommand{\cth}{ \cos\theta }
\newcommand{\sth}{ \sin\theta }
\newcommand{\dOm}{d\Omega}
\newcommand{\vvphi}{\vec{v}_{\phi}}
\newcommand{\demi}{\frac{1}{2}}
\newcommand{\omb}{\overline{\omega}}
\newcommand{\rhobar}{\overline{\rho}}
\newcommand{\ephi}{\vec{e}_\phi}
\newcommand{\ddr}[1]{\frac{{\rm d}  #1}{{\rm d} r}}
\newcommand{\dr}[1]{\frac{\partial  #1}{\partial r}}
\newcommand{\dt}[1]{\frac{\partial  #1}{\partial t}}
\newcommand{\dnt}[1]{\frac{{\rm d}  #1}{{\rm d}t}}
\newcommand{\dnx}[1]{\frac{{\rm d}  #1}{{\rm d}x}}
\newcommand{\dny}[1]{\frac{{\rm d}  #1}{{\rm d}y}}
\newcommand{\dnz}[1]{\frac{{\rm d}  #1}{{\rm d}z}}
\newcommand{\Dt}[1]{\frac{D #1}{D t}}
\newcommand{\dx}[1]{\frac{\partial  #1}{\partial x}}
\newcommand{\dy}[1]{\frac{\partial  #1}{\partial y}}


\bibliographystyle{plain}

\thesaurus{}

\title{Anisotropic diffusion and shear instabilities}
\author{S. Talon \& J.-P. Zahn}
\offprints{S. Talon}

\institute{
D\'epartement d'Astrophysique Stellaire et Galactique, Observatoire de
 Paris, Section de Meudon, 92195 Meudon, France}

\date{Received . . .; accepted . . .}

\maketitle
\markboth{S. Talon \& J.-P. Zahn: Anisotropic diffusion and shear instabilities }{}

\begin{abstract}
We examine the role of anisotropic turbulence on the shear instabilities in
a stratified flow. Such turbulence is expected to occur in the
radiative interiors of stars, due to their differential rotation and their
strong stratification, and the turbulent transport associated with it will
be much stronger in the horizontal than in the vertical direction. It will
thus weaken the restoring force which is caused by the gradient of mean
molecular weight ($\mu$).

We find that the critical shear 
which is able to overcome the $\mu$-gradient is substantially reduced by this
anisotropic turbulence, and we derive an estimate for the resulting
turbulent diffusivity in the vertical direction. 

\end{abstract}

\section{Introduction}
The recent observations performed by Herrero et al. (1992)
reveal significant abundances anomalies in O and B stars.
Even though these observations do not involve a high
number of stars, they display a striking correlation
between the rotation velocity and the helium and nitrogen overabundances. 

Of all instabilities related with the (differential) rotation which
have been examined so far, shear instabilities are the most
efficient. They operate on a dynamical timescale, unlike the GSF instability
(Goldreich \& Schubert 1967 and Fricke 1968) or the ABCD instability
(Knobloch \& Spruit 1983), which both are of double diffusive nature and act
on a thermal, hence much longer, timescale. In stellar radiation zones, the
shear instability is hindered by the stable stratification, but nothing can
prevent it if the flow is sheared in the horizontal direction. For this
reason, as was argued by Zahn (1975), one should expect that the
differential rotation in latitude induced by the meridian circulation
maintains a turbulence which is strongly anisotropic, with a horizontal
component of the turbulent diffusivity ($D_h$) much larger than the vertical
one ($D_v$).

This turbulence tends to reduce the differential rotation which
causes the instability, but it has little effect on the transport in the
vertical direction.  To obtain such a vertical transport, we have to invoke
the instability generated by the vertical shear, although it is impeded by
the stratification. Fortunately, the stabilizing effect of the temperature
stratification is weakened by radiative losses, as was first pointed out by
Townsend (1958), and this mechanism plays an important role in stellar interiors
(Zahn 1974).

However in stars the stratification is
due not only to the temperature gradient, but also to the
variation of the mean molecular weight with depth, in
particular outside the regressing convective core
of massive stars. These molecular weight gradients could well
suppress the shear instability, if it were not for the anisotropic
turbulence mentioned above. The horizontal diffusion associated with it will
diminish the effect of the vertical molecular weight gradient, much like the
thermal diffusivity reduces the effect of temperature stratification. We
shall examine in the present paper how this modifies the classical Richardson
criterion and we shall give an estimate for the vertical diffusion
coefficient.


\section{The effect of thermal diffusion on the shear instability}

The condition for the shear instability to occur may be obtained by
stating that the work done against gravity to exchange two ``bubbles''
at levels $z$ and $z+\delta z$ with velocities $U$ and $U+\delta U$
must be smaller than the kinetic energy released by equalling
their velocities ($cf.$ Chandrasekhar 1961); this yields the
familiar Richardson criterion
\beq 
\frac {g}{\rho} \frac{{\rm d} \Delta\rho / {\rm d}z}{\lp {\rm d} U /
{\rm d}z\rp ^2} < \frac{1}{4} , \label{cond_instab}
\eeq
where $z$ is the vertical coordinate, $g$ the gravity, $\rho$ the
density, and $\Delta\rho$ is the density
difference between the bubble and the exterior. It remains to evaluate
that density gradient ${\rm d}\Delta \rho/{\rm d} z$.

Quite generally, we may write
\beqa
 \frac{{\rm d}\,\Delta \ln\rho}{{\rm d}z} &=& \delta \lp \frac {{\rm d}
\ln T}{{\rm d}z} -\frac {{\rm d} \ln T'}{{\rm d}z} \rp - \varphi \frac
{{\rm d} \ln \mu}{{\rm d}z}  \nonumber \\
&=& \frac {1}{H_P} \lc \delta \lp \nabla '
- \nabla \rp + \varphi \nabla_{\mu} \rc ,
\eeqa
where the $'$ designates the conditions inside of the bubble, 
the
gradients being expressed from now on with respect to the pressure,
as is normally done in stellar structure theory:
\beq
 \nabla = \frac {{\rm d} \, \ln T}{{\rm d}\, \ln P} 
\hspace{0.5cm} {\rm and} \hspace{0.5cm}
 \nabla _{\mu} = \frac {{\rm d}\, \ln \mu}{{\rm d}\, \ln P}.
\nonumber  
\eeq
We take the usual notations for
\beq
\delta = -\lp \frac{\partial \ln\rho}{\partial \ln T}\rp _{P,\mu} 
\hspace{0.5cm} {\rm and} \hspace{0.5cm}  
\varphi= \lp \frac{\partial \ln \rho}{\partial \ln \mu}\rp _{P,T}
\nonumber  \eeq
which depend on the equation of state ($\delta =1,
\varphi = 1$ for a perfect gas).

In the absence of thermal diffusion, $\nabla '$ is equal to the
adiabatic gradient $\nabla _{\rm ad}$ and one retrieves the
Richardson criterion relevant for a compressible fluid:
\beq
\frac{N^2}{\lp {\rm d} U /{\rm d}z\rp ^2} < Ri_c,
\hspace{0.5cm} {\rm where} \hspace{0.5cm}
N^2 = N^2_T + N^2_\mu ,
\label{Ri_dyn}
\eeq 
having split the Brunt-V\"{a}is\"{a}l\"{a} frequency into 
 \beq
N^2_T = \frac{g \delta}{H_P} \lp \nabla _{\rm ad} - \nabla \rp
\hspace{0.5cm} {\rm and} \hspace{0.5cm}
N^2_\mu = \frac{g\varphi}{H_P}   \nabla _\mu ,
\eeq
with $Ri_c
\approx 1/4$ being the critical Richardson number.

The effect of thermal damping has been examined by Dudis (1974), 
and Zahn (1974) has given an estimate for the turbulent diffusivity in
the limit of low P\'eclet number ($v \ell /K \ll 1$, with $K$
being the thermal diffusivity). Recently, Maeder
(1995) derived a new expression for this coefficient, which has the
advantage of describing the intermediate regimes.
It is inspired by
the mixing length treatment of non-adiabatic convection (see Cox \&
Giuli 1968), where one defines the ratio $\Gamma$ between the thermal
energy transported by an eddy  of volume $\cal V$
\beq
\rho {\cal V} C_p  \Delta T  = \rho {\cal V}  C_p T \frac{\ell}{H_P}
(\nabla' - \nabla)
\eeq
and the energy which diffuses through its surface of area $\cal A$
during its lifetime $\ell/v$ 
\beqan
{\cal A} \chi \lp \frac{\Delta T/2}{\ell/2} \rp \frac{\ell}{v} &=&
\rho {\cal V} T \Delta S \\
&=& \rho {\cal V} C_P T \frac{\ell}{H_P} (\nabla_{\rm ad} -
\nabla') \nonumber
\eeqan
($\Delta T$ is the maximum temperature difference, $\Delta S$ is 
the entropy excess, and 
$\chi = \rho C_P K = 16 \sigma T^3 / 3 \kappa \rho$  the radiative
conductivity, with the usual notations). Assuming that the eddy may be
described as a sphere of diameter $\ell$ ($i.e. \; {\cal V}/ {\cal A} = \ell
/6$),  one has  \beq 
\Gamma = \frac{v \ell}{6 K} =
\frac{\nabla' - \nabla}{\nabla_{\rm ad} - \nabla'}.
\eeq

This yields the modified version of the Richardson criterion which was
established by Maeder (1995):   
\beq  
g \frac{{\rm d}\Delta \ln \rho}{{\rm d}z} =  \lp \frac {\Gamma}{\Gamma +1}
\rp N^2_T + N^2_\mu  \leq Ri_c \lp \frac{{\rm d} U}{{\rm d}z}\rp ^2 .
\label{Rich_maeder}  
\eeq
According to this criterion, the vertical shear can be unstable only provided
 \beq
 \lp \frac {{\rm d}U}{{\rm d}z} \rp ^2 >
\frac{1}{Ri_c} \, N^2_\mu.
\label{Ri_mu}
\eeq
Then eq.(\ref{Rich_maeder}) may be solved for the parameter $\Gamma =
v\ell/6K$ which characterizes the largest eddies that satisfy the
Richardson criterion, and these produce a turbulent diffusivity 
$D_v = 2 K \Gamma$.

Recently Meynet \& Maeder (1996) calculated the evolution of rotating
massive stars with this prescription for the
turbulent diffusion, assuming local conservation of angular momentum. They
found that, even though the model displays a strong
differential rotation at the end of the main sequence,
no significant mixing occurs because the molecular weight gradient
inhibits the shear instability. Since according to the observations 
such mixing does take place in fast rotators, they conclude
that this prescription underestimates the shear mixing in stars, and there
must be some way to overcome the $\mu$-barrier.

\section{The effect of the horizontal diffusion}

As was already mentioned in the Introduction, it seems plausible that 
stars exhibit an anisotropic turbulence due to the horizontal shear which
is induced by the meridian circulation (see also Zahn 1992). This 
turbulence generates a horizontal transport which is much larger than the
vertical one, and which tends to smooth out the horizontal fluctuations of
molecular weight. Therefore $\mu$ will vary along the trajectory of the
bubble considered above (see fig. 1).

The restoring force per unit mass is then given by
\beq 
 g \frac{{\rm d}\Delta \ln\rho}{{\rm d}z} =  \frac{g}{H_P} \lc \delta
\lp \nabla ' - \nabla \rp - \varphi \lp \nabla _{\mu}' - \nabla _{\mu}
\rp \rc, \label{Rich_mu}
\eeq
with the same notations as before.

\begin{figure}     %
\hspace{0.5cm}
\psfig{figure=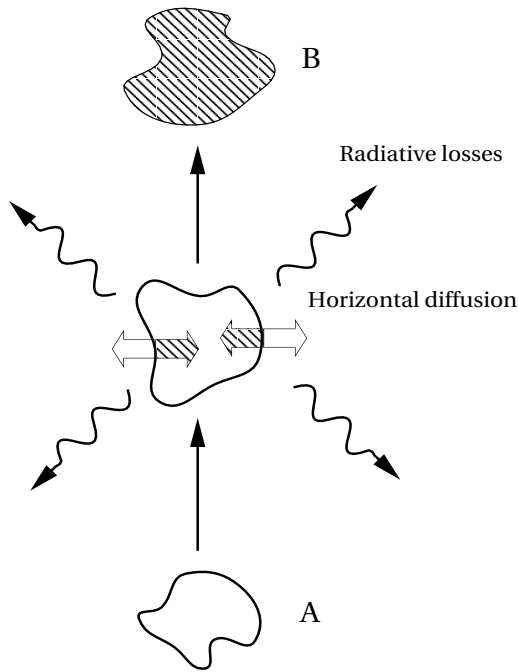,height=8.8cm,width=6.8cm}
\caption[]{As a turbulent eddy moves vertically, it is partially mixed
with the surrounding medium, and therefore its composition varies.
}
\end{figure}

To estimate $(\nabla _{\mu}' - \nabla _{\mu})$ we proceed as above
for the thermal diffusion of heat: it suffices to replace $\Delta T$
by $\Delta \mu$ and $K$ by the turbulent diffusivity $D_h$. The result
is
\beq
\nabla _{\mu}' - \nabla _{\mu}  = 
- \frac {\Gamma _{\mu}}{\Gamma _{\mu}+1}  \nabla _{\mu},
\eeq
where now $\Gamma _{\mu}= v\ell/6 D_h$.

We thereby obtain still another version of the Richardson criterion:
\beq  
 \lp \frac {\Gamma}{\Gamma +1} \rp N^2_T 
+ \lp \frac {\Gamma_\mu}{\Gamma_\mu +1} \rp N^2_\mu 
\leq Ri_c \lp \frac{{\rm d} U }{{\rm d}z}\rp ^2 ,
\label{Rich_TZ}  
\eeq
which describes how the stabilizing effect of the $\mu$-gradient is
diminished by the horizontal turbulence.
Note that the influence of the temperature gradient will also be modified by
this turbulence since the advective transport of heat will be
added to the radiative diffusion: $K \rightarrow (K+D_h)$.

We see that any vertical shear may become unstable, 
for small enough $v \ell$.  But the Reynolds number characterizing these
eddies must be larger than some critical value, otherwise their motion will
be damped out through viscous friction. Hence they must obey the condition 
\beq
\frac{v \ell}{\nu} \le  Re_c,
\label{Reynolds}
\eeq
where $\nu$ is the molecular viscosity, and $Re_c \simeq 10$ is the
critical Reynolds number.

Among all turbulent eddies satisfying (\ref{Rich_TZ}) and
(\ref{Reynolds}), those
contributing most to the vertical transport are those for which $v\ell$ is
the largest. Their size will be given by the equality in
eq.~(\ref{Rich_TZ}), namely  
\beq
\frac{x}{x+K+D_h} N^2_T + \frac{x}{x+D_h} N^2_{\mu} = Ri_c \lp
\frac {{\rm d}U}{{\rm d}z} \rp ^2 \label{sec_ordre}
\eeq
where $x=v \ell/6 \ge \nu Re_c/6$.

It is an easy matter to solve this second order equation
(\ref{sec_ordre}), and to derive from it the value of vertical diffusivity
$D_v \simeq v \ell /3$. Solutions exist only if
\beq
 \lp \frac {{\rm d}U}{{\rm d}z} \rp ^2 \geq
\frac{\nu Re_c}{6 Ri_c} \lc\frac{N^2_T}{K+D_h} + \frac{N^2_\mu}{D_h}\rc,
\eeq
which replaces the much more stringent condition (\ref{Ri_mu}).

Note that the vertical turbulence transports heat and thus modifies the
temperature gradient $\nabla$. This was taken into account by Maeder \& Meynet
(1996) when they evaluated the mixing by shears.
However, in most circumstances relevant to radiative interiors,
we are in the presence of {\it mild turbulence},
so that  $x \ll K$, and we may ignore this effect. Furthermore, we stated
as a working hypothesis the condition $D_v \simeq 2x \ll D_h$.
The turbulent diffusivity is then given by the simple expression
\beq
D_v \simeq \frac{2 Ri_c ({\rm d}U/{\rm d}z)^2} {N^2_T/(K+D_h) + N^2_{\mu}/D_h}.
 \label{TZ}
\eeq

One final word about the value of the ratio ${\cal{V}}/{\cal{A}}$
between volume and area of the turbulent eddies. Its choice is rather
arbitrary. Following Maeder (1995), we have taken ${\cal{V}}/{\cal{A}} =
\ell/6$, whereas B\"{o}hm-Vitense (1958) proposed the value 
${\cal{V}}/{\cal{A}} = (2/9) \ell$  when she developed her famous mixing
length formalism.  
We could refine this numerical coefficient, for instance by taking our
inspiration from Dudis' (1974) work, where $\Gamma = (3/8) v\ell/K$.
However, we prefer to hide this uncertainty 
in the choice of $Ri_c$ and in the prescription we use to evaluate the
horizontal diffusion coefficient $D_h$, which also involves an unknown
parameter of order unity (see Zahn 1992).

\section{Conclusions}

We have shown how the Richardson criterion, which governs the
shear instabilities in a stratified fluid, is modified in the presence of
a strongly anisotropic turbulence. This turbulent erosion is similar to
that which interfers with the advective transport of a large-scale flow, and
which has been described by Chaboyer and Zahn (1992). 
Another example of this phenomenon is given in a numerical simulation
performed by Vincent et al. (1996). Here it diminishes the
stabilizing action of a molecular weight gradient, to a point which
makes it plausible that rapidly rotating stars be partially
mixed. The prescription (\ref{TZ}) we have derived above for the vertical
diffusivity is being implemented in our stellar evolution codes, and the
results will be reported in a forthcoming paper.

\begin{acknowledgements}
We would like to thank A. Maeder and G. Meynet for the interesting discussions
we had with them on the subject.
S.T. gratefully acknowledges support from NSERC of Canada.
\end{acknowledgements}

\end{document}